\documentclass[conference]{IEEEtran}
\IEEEoverridecommandlockouts
\usepackage{Shortcuts}
\def\BibTeX{{\rm B\kern-.05em{\sc i\kern-.025em b}\kern-.08em
    T\kern-.1667em\lower.7ex\hbox{E}\kern-.125emX}}
\begin{document}

\title{"Machine LLRning":\\Learning to Softly Demodulate
}

\author{\IEEEauthorblockN{Ori Shental}
\IEEEauthorblockA{\comment{\textit{Communications Theory Department} \\}
\textit{Bell Labs}\\
Holmdel, New Jersey 07733, USA \\
ori.shental@nokia-bell-labs.com}
\and
\IEEEauthorblockN{Jakob Hoydis}
\IEEEauthorblockA{\comment{\textit{Hardware Platforms \& Research Testbeds Department} \\}
\textit{Bell Labs}\\
91620 Nozay, France \\
jakob.hoydis@nokia-bell-labs.com}
}

\maketitle

\begin{abstract}
Soft demodulation, or demapping, of received symbols back into their conveyed soft bits, or bit log-likelihood ratios (LLRs), is at the very heart of any modern receiver. In this paper, a trainable universal neural network-based demodulator architecture, dubbed "LLRnet", is introduced. LLRnet facilitates an improved performance with significantly reduced overall computational complexity. For instance for the commonly used quadrature amplitude modulation (QAM), LLRnet demonstrates LLR estimates approaching the optimal log maximum a-posteriori inference with an order of magnitude less operations than that of the straightforward exact implementation. Link-level simulation examples for the application of LLRnet to 5G-NR and DVB-S.2 are provided. LLRnet is a (yet another) powerful example for the usefulness of applying machine learning to physical layer design.
\end{abstract}


\section{Introduction}
The basic procedure of demapping received symbols back into their embedded soft bits, feeding a following stage of error correction decoding, is a crucial component in any modern communication system. The soft bit, quantifying the level of confidence in the quality of the symbol's demodulation process, is typically expressed in terms of the \emph{log-likelihood ratio} (LLR).

An exact evaluation of the LLR is achieved via a scheme, known as the \emph{log-MAP}~\cite{bib:Erfanian1994}, computing the logarithm of the ratio between the maximum a-posteriori (MAP) probabilities of the latent bit's two hypotheses given the observed symbol. Though statistically optimal, the computational complexity of the log-MAP algorithm scales with the size of the symbol constellation, making its direct implementation impractical in realistic systems.

A popular approximation of the optimal log-MAP rule, serving as a feasible golden target in designing practical systems, is the well-known \emph{max-log-MAP} algorithm~\cite{bib:Robertson1995}. However, although eliminating the need to compute complex exponential and logarithmic functions, in general the approximated max-log-MAP algorithm still consumes, on a symbol-rate basis, an extensive number of operations which scales with the modulation order.

There are certain constellation schemes, like the Gray-coded quadrature amplitude modulation (QAM), for which the max-log-MAP boils down to a reduced-complexity piecewise linear function and can be thus implemented, for instance, via a lookup table (LUT)~\cite{bib:Tosato2002}. This simplification is attributed to the QAM's inherent constellation mapping symmetries and its separability into two independent pulse amplitude modulations (PAM) in the real and imaginary parts of the symbol. As shall be exemplified in the sequel, although the computational ease in the operation of the max-log-MAP algorithm on QAM constellations, it still suffers from significant performance degradation in the estimation of the LLRs compared to the exact log-MAP, especially in the low and intermediate signal-to-noise ratio (SNR) regimes. This degradation is especially detrimental in the (almost) inevitable pragmatic case of non-ideal link adaptation. Further computationally simplified versions of the max-log-MAP for either QAM~\cite{bib:Tosato2002} or other modulation schemes, like phase-shift keying (PSK)~(\eg, \cite{bib:Wang2014} and references therein), suffer from an additional performance penalty w.r.t. the original "full-blown" max-log-MAP algorithm.

In this contribution, a machine-learning architecture for efficient universal soft demodulation, dubbed \emph{"LLRnet"}, is proposed. One should bear in mind that the demapping procedure can be simply abstracted as nothing but a symbol-to-bit LLRs function. Now, neural networks are well-known as a great tool in effectively approximating functions (viz. Cybenko's universal approximation theorem~\cite{bib:Cybenko1989}). Hence it seems very natural and beneficial to train a neural network to directly learn the functionality of either the impractical exact log-MAP, the cumbersome max-log-MAP, or any other expert-based target demapping rule. As shall be shown, the LLRnet provides an excellent mechanism for achieving the target demodulation rule performance. For the family of QAM constellations, LLRnet is shown to practically reproduce the exact log-MAP algorithm with a substantially reduced computational burden (\eg, ten times less the operations for 1024-QAM).

The paper is organized as follows. The LLR estimation problem is described in Section~\ref{sec_problem}. Section~\ref{sec_LLRnet}
introduces the proposed LLRnet architecture for soft demodulation. Section~\ref{sec_Examples} provides simulation results and discusses the performance of LLRnet in the context of two commercial system applications, namely the cellular 5G and the satellite DVB-S.2 standards. Finally, Section~\ref{sec_Conclusion} contains some concluding remarks.

\section{Problem Formulation}\label{sec_problem}
\begin{figure}[t!]
    \centering
    \includegraphics[width=\columnwidth,bb=0 0 380 96]{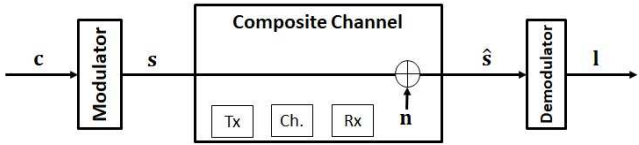}
    \caption{A communication system comprising transmit modulator, composite channel and receive demodulator. The composite channel encapsulates arbitrary processing stages of succeeding transmitter, physical channel, and receiver.}
    \label{fig_system_model}
\end{figure}

Consider a generic communication system (as depicted in Fig.~\ref{fig_system_model}) with a modulator at the transmitter mapping an incoming stream of encoded bits (where each bit is denoted by $c\in\{0,1\}$) to an outgoing stream of modulated symbols, $\vs\triangleq\{s_{1},\ldots,s_{N}\}^{T}\in\mathcal{C}$, chosen from an arbitrary finite set of constellation points in a (possibly hyper-dimensional) complex domain, $\mathcal{C}\subset\mathbb{C}^{N}$, $N\in\mathbb Z_{> 0}$. To this end, let $\vc\triangleq\{c_{1},\ldots,c_{M}\}^{T}\in\{0,1\}^{M}$, be a vector of some arbitrary $M$ consecutive bits in the stream being modulated to an $N$-dimensional complex symbol $\vs$.\comment{\footnote{Note that in most commercial systems $N=1$ is used.}}\comment{It is assumed that the modulator's output power is normalized, namely $\norm{\vs}_{2}=1$.} Hereinafter, it is assumed that all possible $M$-bit vectors, $\vc$, are equiprobable, thus all the modulated symbols in the constellation set $\mathcal{C}$ are equally likely to be transmitted. Incorporating knowledge, \eg, iteratively originated from the error correction decoding stage, on (unequal) prior probabilities of the modulated symbols in the discussed demodulation schemes is straightforward.

The symbol $\vs$ is transmitted through a composite channel comprising, in addition to the physical channel, an abstraction of other critical transceiver processing steps. On the transmitter side, these may consist of (but not limited to) precoding for transmission over multiple-input multiple-output (MIMO) systems, mapping to orthogonal frequency-division multiplexing (OFDM), digital-to-analog conversion (DAC) and analog mixing on a radio carrier. On the receiver side, the composite channel may encapsulate, \eg, analog-to-digital conversion (ADC), down-mixing to baseband, synchronization, filtering, OFDM demodulation, MIMO combining and channel estimation and equalization. These transceiver processing steps are standard and do not explicitly pertain to the problem under study of demapping, or soft demodulation, of the received symbol to bit LLRs.

The demodulator at the receiver observes the composite channel's complex-valued symbol estimate, $\hat{\vs}\in\mathbb{C}^{N}$, where the effect of the composite channel is captured by the relation between the transmitted and observed symbols, $\vs$ and $\hat{\vs}$, respectively. Although not necessarily accurate, for the purpose of soft demodulation to bit LLRs, the composite channel is typically modeled as an additive white Gaussian noise (AWGN) channel, to yield
\BE\label{eq_model}
    \hat{\vs}=\vs+\vn,
\EE
where the $N$-dimensional AWGN vector $\vn\sim\mathcal{CN}(\mathbf{0},\sigma^{2}\mI_{N})$ and $\sigma^{2}\geq\sigma_{0}^{2}$. The scalar  $\sigma_{0}$ denotes the standard deviation of the ambient noise in the physical channel, corresponding to some SNR. It is important to note that this work does not focus on optimizing the, so called, "composite channel" itself (\eg, via improving MIMO demodulation), but in streamlining symbol demapping by adopting a machine-learning approach. To this end, the observed symbol estimate, $\hat{\vs}$, can be viewed, for instance, as an equalized symbol following a linear detection stage.

Now mainly for the purpose of facilitating an improved decoding in a succeeding stage, the demodulator demaps the observed symbol, $\hat{\vs}$, into a vector of estimates of the $M$-bit LLRs, $\vl\in\mathbb{R}^{M}$, corresponding to the confidence in the inference of each of the original coded bits in $\vc$. The $i$'th entry of the LLRs vector, $l_{i}$, adheres to the logarithm of the MAP relation
\BE\label{eq_relation}
    l_{i}\triangleq\log\Bigg(\frac{\Pr{(c_{i}=0|\hat{\vs})}}{\Pr{(c_{i}=1|\hat{\vs})}}\Bigg),\quad i=1,\ldots,M.
\EE
An \emph{exact} computation of the log-MAP expression~\eqref{eq_relation}, under the examined model~\eqref{eq_model}, yields
\BE\label{eq_logMAP}
    l_{i}=\log\frac{{\sum_{\vs\in\mathcal{C}_{i}^{0}}\exp{\Big(-\frac{\norm{\hat{\vs}-\vs}_{2}^{2}}{\sigma^{2}}\Big)}}}{\sum_{\vs\in\mathcal{C}_{i}^{1}}\exp{\Big(-\frac{\norm{\hat{\vs}-\vs}_{2}^{2}}{\sigma^{2}}\Big)}},\quad i=1,\ldots,M,
\EE
where $\mathcal{C}_{i}^{\delta}\in\mathcal{C}$ is the subset of constellation points, known at the receiver side, for which the $i$'th bit is equal to $\delta\in\{0,1\}$.
Applying the approximation
\BE
    \log\Bigg(\sum_{j}\exp\big(-x_{j}^{2}\big)\Bigg)\approx\max_{j}\big(-x_{j}^{2}\big) \nonumber
\EE
on the exact log-MAP operation~\eqref{eq_logMAP}, \emph{approximated} LLR estimates can be derived from a simplified rule, well-known as the max-log-MAP
\BE\label{eq_maxlogMAP}
    l_{i}\approx\frac{1}{\sigma^{2}}\Bigg(\min_{\vs\in\mathcal{C}_{i}^{1}}\norm{\hat{\vs}-\vs}_{2}^{2}-\min_{\vs\in\mathcal{C}_{i}^{0}}\norm{\hat{\vs}-\vs}_{2}^{2}\Bigg),\quad i=1,\ldots,M.
\EE

The computational complexity per received symbol of a brute-force implementation of either the log-MAP~\eqref{eq_logMAP} or the max-log-MAP~\eqref{eq_maxlogMAP} algorithms scales with the size of the constellation, $|\mathcal{C}|$, to yield $\mathcal{O}(2^{MN})$ operations. However, the approximated max-log-MAP demapping rule has the advantageous property of eliminating the need for computing complex exponential and logarithmic functions. For dealing with the popular (Gray-coded) QAM constellations, although being severely sub-optimal the max-log-MAP is typically the common demapper of choice due to the fact that it can be reduced to a single LUT implementation of piecewise linear functions, which scales linearly with SNR. Note that the exact log-MAP algorithm can be only crudely approximated via multiple LUTs, essentially one for each SNR working point~\cite{bib:Yao2015}.

\section{Soft Demodulation with LLRnet}\label{sec_LLRnet}
\begin{figure}[t!]
    \centering
    \includegraphics[width=\columnwidth,bb=0 0 509 360]{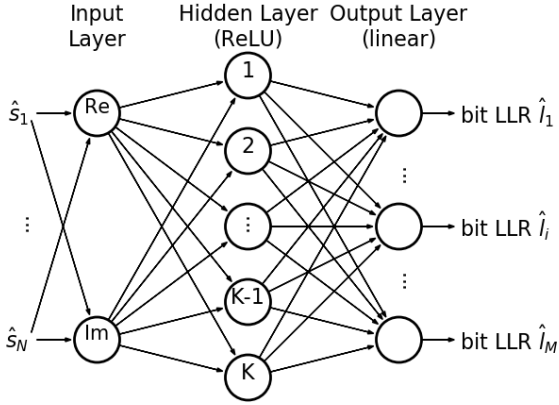}
    \caption{LLRnet architecture composed of: An input of $N$-dimensional received symbol split into its real and imaginary parts; $K$-neurons hidden layer with (\eg) ReLU activation function; An $M$-bit LLR linear output layer.}
    \label{fig_model}
\end{figure}

LLRnet is a neural network designed to learn a desired soft demodulation scheme, demapping a complex symbol to its conveyed bits' real-valued LLRs. The architecture of LLRnet is illustrated in Fig.~\ref{fig_model}. The input layer of the LLRnet is fed by the received symbol estimate vector, $\hat{\vs}$, where it is divided into its real and imaginary parts. The output vectors of the two input nodes are injected to a hidden layer of $K$ neurons. The scalar output of the $k$'th neuron is determined by $y_{k}=f(\vw_{k}^{T}\vx_{k}+b_{k})$, where $\vx_{k},\vw_{k}\in\mathcal{R}^{2N}$ and $b_{k}\in\mathcal{R}$ are the $k$'th neuron's input vector, weights vector and bias, respectively. The operator $f(\cdot)$ denotes the neuron's transfer, or activation, function. Unless otherwise stated, in its hidden layer LLRnet uses a rectified linear unit (ReLU), namely $f(x)=x^{+}=\max(0,x)$. The outputs of the $K$ hidden neurons are then inserted to a linear output layer of $M$ nodes. The $m$'th output of the LLRnet, estimating the $m$'th bit LLR, is determined by $\hat{l}_{m}=\boldsymbol{\omega}_{m}^{T}\vy+\beta_{m}$, where $\vy\triangleq\{y_{1},\ldots,y_{K}\}^{T}\in\mathcal{R}^{K}$ is a concatenation of the hidden layer outputs, while $\boldsymbol{\omega}_{m}\in\mathcal{R}^{K}$ and $\beta_{m}\in\mathcal{R}$ are the $m$'th output node's weights vector and bias, respectively. The set of all trainable parameters of the LLRnet, namely all the weights $\vw_{k}$, $\boldsymbol{\omega}_{m}$ and biases $b_{k}$, $\beta_{m}$, are denoted by $\boldsymbol{\theta}$.

\begin{figure}[t!]
    \centering
    \includegraphics[width=\columnwidth,bb=0 0 217 219]{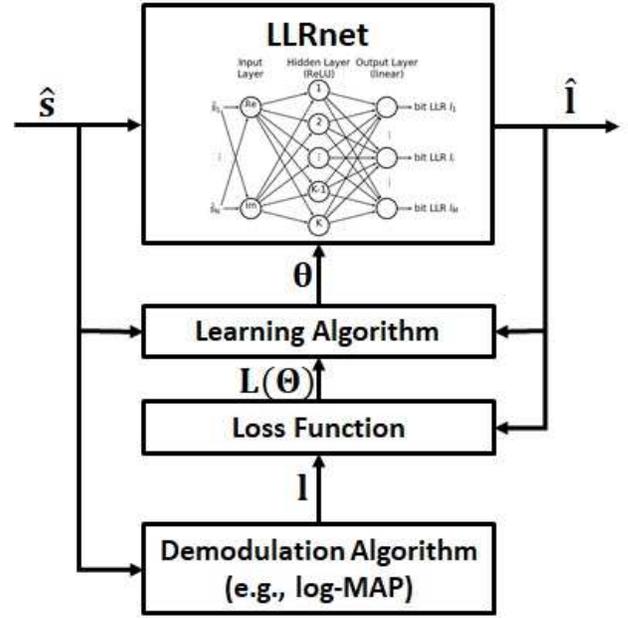}
    \caption{Block diagram for the training of LLRnet.}
    \label{fig_diagram}
\end{figure}

Fig.~\ref{fig_diagram} schematically describes the training process of the neural demodulator. Since the LLRnet's input-to-output function is fully differentiable w.r.t. the set of all trainable parameters, $\boldsymbol{\theta}$, a gradient-based training approach can be adopted. The trainable parameters are first randomly initialized. For a certain batch of $B$ received symbols, $\hat{\vs}^{(b)}$, $b=1,\ldots,B$, compute the corresponding bit LLRs via a desired demapping algorithm, \eg~log-MAP or max-log-MAP (for implementing these specific target demodulation algorithms, an estimate of $\sigma^{2}$ is also required).\footnote{Any other expert-based demapping rule can be targeted, \eg, in the case of colored noise due to equalization.} Then feed both the conventionally computed LLRs, $\vl^{(b)}$, along with the LLRs computed through LLRnet, $\hat{\vl}^{(b)}$, into a loss function which is evidently a function of the trainable set of parameters $\boldsymbol{\theta}$. Such a loss function could be, for instance, the mean-squared error (MSE)
\BE
    L^{\text{MSE}}(\boldsymbol{\theta})\triangleq\frac{1}{B}\sum_{b=1}^{B}\norm{\hat{\vl}^{(b)}-\vl^{(b)}}_{2}^{2}, \nonumber
\EE
or alternatively the cross-entropy function
\BE
    L^{\text{CE}}(\boldsymbol{\theta})\triangleq-\frac{1}{B}\sum_{b=1}^{B}\sum_{m=1}^{M}l_{m}^{(b)}\log(\hat{l}_{m}^{(b)})+(1-l_{m}^{(b)})\log(1-\hat{l}_{m}^{(b)}). \nonumber
\EE
Define a stop criterion which can be either a fixed number of iterations, a threshold on the loss or a number of iterations during which the loss
has not decreased. Unless the stop criterion is met, update the parameter set, $\boldsymbol{\theta}$, based on a learning algorithm using gradient descent,
\BE
    \boldsymbol{\theta}\leftarrow\boldsymbol{\theta}-\alpha\nabla_{\boldsymbol{\theta}}L(\boldsymbol{\theta}), \nonumber
\EE
where $\alpha>0$ is the learning rate. Compute the loss function again under the newly learned set of trained parameters. When the stop criterion is met and training is complete, one moves to the inference stage where the LLRnet serves as the sole demodulator, efficiently imitating the functionality of the desired soft demodulator. In the following section LLRnet is simulated and its performance is evaluated in two end-to-end system usecases.

\section{Simulation Examples}\label{sec_Examples}
\subsection{Throughput of PDSCH in 5G-NR}\label{sec_5G}

\begin{figure}[hp]
    \centering
    \begin{subfigure}[b]{\columnwidth}
        \centering
        \includegraphics[width=\columnwidth,bb=0 0 225 172]{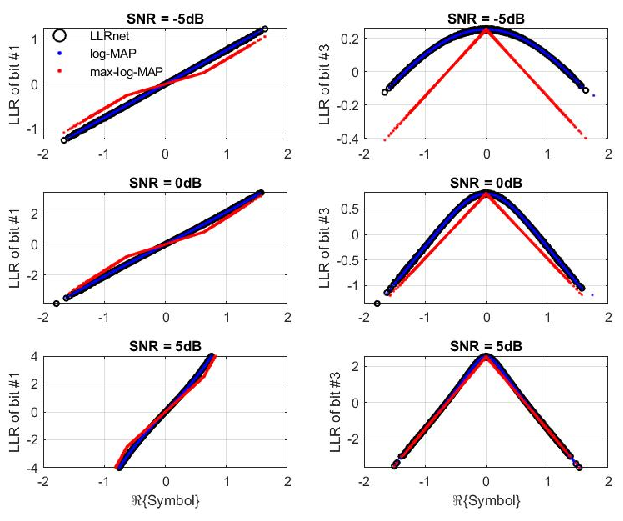}
        \caption{16-QAM}
    \end{subfigure}
    \vskip\baselineskip
    \begin{subfigure}[b]{\columnwidth}
        \centering
        \includegraphics[width=\columnwidth,bb=0 0 225 172]{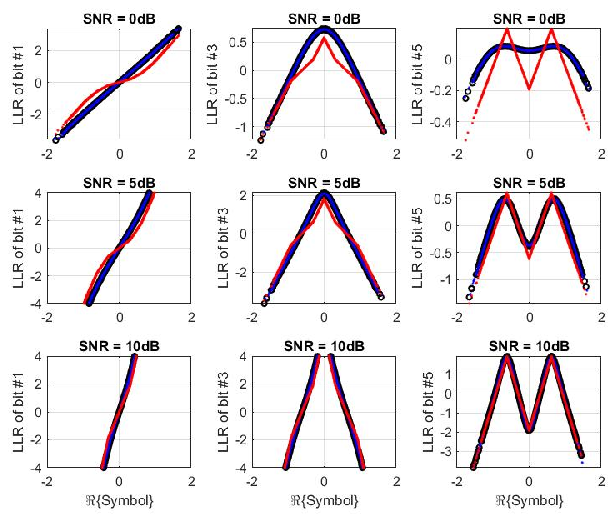}
        \caption{64-QAM (Legend is identical to that of (a).)}
    \end{subfigure}
    \vskip\baselineskip
    \begin{subfigure}[b]{\columnwidth}
        \centering
        \includegraphics[width=\columnwidth,bb=0 0 225 172]{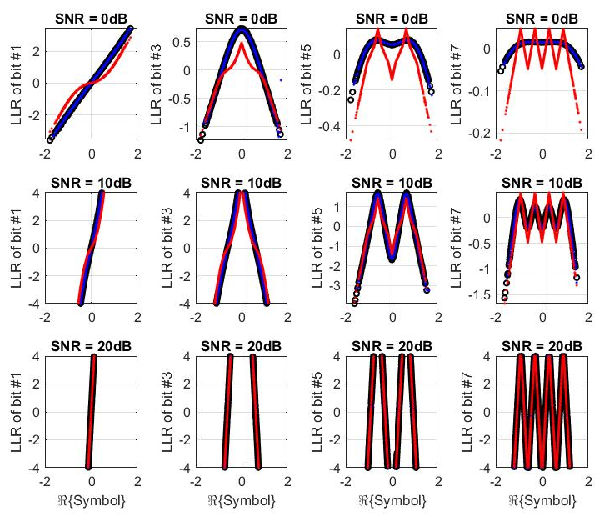}
        \caption{256-QAM (Legend is identical to that of (a).)}
    \end{subfigure}
    \caption{LLR, derived via log-MAP, max-log-MAP and LLRnet, as a function of the real part of the symbol for the odd bits in 16/64/256-QAM.}
    \label{fig_LLR_QAM}
\end{figure}

%
%

\begin{figure}[hp]
    \centering
    \begin{subfigure}[b]{\columnwidth}
        \centering
        \includegraphics[width=\columnwidth,bb=0 0 149 115]{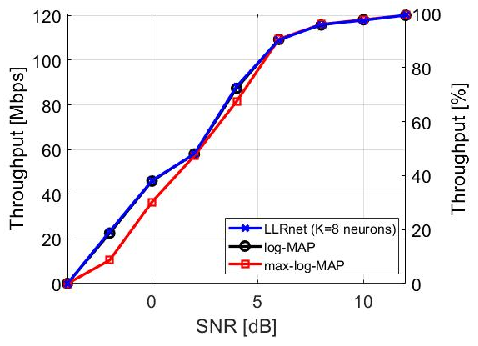}
        \caption{16-QAM}
    \end{subfigure}
    \vskip\baselineskip
    \begin{subfigure}[b]{\columnwidth}
        \centering
        \includegraphics[width=\columnwidth,bb=0 0 149 115]{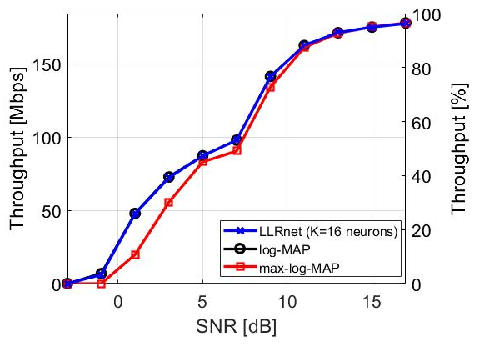}
        \caption{64-QAM}
    \end{subfigure}
    \vskip\baselineskip
    \begin{subfigure}[b]{\columnwidth}
        \centering
        \includegraphics[width=\columnwidth,bb=0 0 149 115]{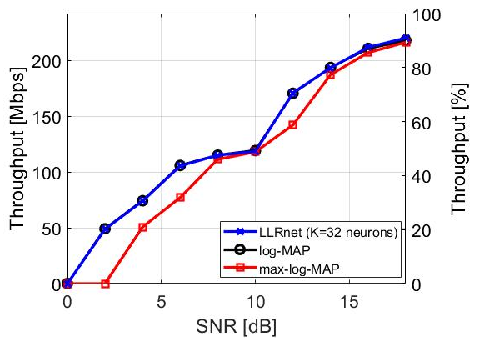}
        \caption{256-QAM}
    \end{subfigure}
    \caption{PDSCH throughput, derived via log-MAP, max-log-MAP and LLRnet, as a function of SNR for 16/64/256-QAM.}
    \label{fig_tput}
\end{figure}

%
%

\setlength{\aboverulesep}{0pt}
\setlength{\belowrulesep}{0pt}
\setlength{\extrarowheight}{.75ex}

\begin{table*}[tp]
\captionsetup{size=normalsize,
    skip=5pt, position = bottom}
\renewcommand\thetable{II}
\caption{Computational complexity comparison of the demodulation schemes in terms of real operations per received symbol.}
\label{table_complexity}
    \centering
    \begin{threeparttable}
    \begin{tabular}{laaccccaa}
    \toprule
     & \multicolumn{2}{c}{16-QAM}& &\multicolumn{2}{c}{64-QAM}& &\multicolumn{2}{c}{256-QAM}\\
              \cmidrule{2-3} \cmidrule{5-6} \cmidrule{8-9}
            & {log-MAP} & {LLRnet}
            & & {log-MAP} & {LLRnet}
            & & {log-MAP} & {LLRnet\tnote{*}}\\
            & &  {(K=8)}&
            & &  {(K=16)}&
            & &  {(K=32)} \\
    \midrule
    Multiplication \& division             & 52  & 40&& 198  & 112 && 776 & 288 (320)\\
    Addition \& subtraction            & 104  & 40 && 564  & 112 && 2800 & 288 (352)\\
    Exponent \& logarithm      & 20 & 0 && 70  & 0  && 264 & 0 (64)\\
    Comparator      & 0 & 8 && 0  & 16  && 0 & 32 (0)\\
    \midrule\midrule
    Total & 176 & 88 && 832  & 240 && 3840 & 608 (736)\\
    \bottomrule
    \end{tabular}
    \begin{tablenotes}
        \item[*] Parentheses correspond to using $\tanh$, rather than ReLU, activation function at the hidden layer.
    \end{tablenotes}
    \end{threeparttable}
\end{table*}

\begin{table}[bp]
\renewcommand\thetable{III}
\caption{Computational complexity for 1024-QAM.}
\label{table_complexity_1024}
    \centering
    \begin{tabular}{lcc}
    \toprule
     & \multicolumn{2}{c}{1024-QAM} \\
              \cmidrule{2-3}
            & {log-MAP} & {LLRnet} \\
            & &  {(K=64)}\\
    \midrule
    Multiplication \& division             & 3082  & 704 \\
    Addition \& subtraction            & 13292  & 704 \\
    Exponent \& logarithm      & 1034 & 0 \\
    Comparator      & 0 & 64 \\
    \midrule\midrule
    Total & 17408 & 1472 \\
    \bottomrule
    \end{tabular}
\end{table}

\begin{table}[tp]
\renewcommand\thetable{I}
\caption{Simulated 5G-NR key configuration parameters.}
\label{tab_param}
    \centering
\begin{tabular}{|l|c|}
  \hline
  \# of 10ms frames & 200 \\
  Bandwidth & 20MHz \\
  \# of (12 sub-carriers) resource blocks & 51 \\
  Subcarrier spacing & 30KHz \\
  Cyclic prefix & Normal \\
  \# of Tx. antennas & 8 \\
  \# of Rx. antennas & 2 \\
  \# of layers & 2 \\
  Transport channel coding & LDPC \\
  Target code rate & 0.4785 \\
  Modulation & 16-QAM, 64-QAM, 256-QAM \\
  PDSCH precoding & SVD, single matrix \\
  Waveform & CP-OFDM \\
  Channel model & Clustered delay line (CDL) \\
  Channel estimation & Ideal \\
  MIMO equalization & Linear MMSE \\
  Synchronization & Perfect \\
  HARQ & Enabled, 16 processes \\
  \hline
\end{tabular}
\end{table}

The proposed LLRnet demodulation engine is utilized in the decoding of the physical downlink shared channel (PDSCH) in a 5G New Radio (NR) link, as defined by the 3GPP NR standard. The key configuration parameters of the simulated link are listed in Table~\ref{tab_param}.
In the simulated configuration, a frame (10ms) is divided into 10 subframes, where each subframe (1ms) is composed of 2 slots, and each slot (0.5ms) consists of 14 symbols. For each slot, out of the $51\times12\times14\times2=17136$ resource elements in the received resource grid, $14712$ convey PDSCH symbols.

The training of the LLRnet relies on only about $1\%$ ($148$ symbols) of the $14712$ received PDSCH symbols, taken evenly across a single slot which is, in the reported results, the first slot in a 2 seconds transmission per evaluated SNR point. These $148$ symbols are randomly divided to three sets: (1) about $70\%$ of them are used for training, (2) about $15\%$ used for validation that the trained network has low enough generalization error and for avoiding excessive training which may result in undesirable overfitting, and (3) lastly, about $15\%$ of the symbols are used for a completely independent testing of the LLRnet generalization. Note that the relatively small size of the data set, found sufficient for training the LLRnet, stems from the inherent smoothness of the demapping functions from symbols to bit LLRs.

The chosen training algorithm in this example is the Levenberg-Marquardt backpropagation algorithm~\cite{bib:Hagan1994} and the training process continues until the validation error fails to decrease for $6$ consecutive iterations. The chosen loss function is the MSE, where the desired, or target, demodulation algorithm the LLRnet is trained to reproduce is the exact, yet costly, log-MAP scheme (with $\sigma^{2}=\sigma_{0}^{2}$). The number of neurons in the hidden layer is set to $K=8,16,32$ for the constellations $16,64,256$-QAM (thus $N=1$), respectively. As mentioned previously, the neuron's activation function is a ReLU. Upon completion of the training stage, PDSCH symbols in the consecutive slots are demodulated exclusively by the LLRnet. Re-training, so the LLRnet can readapt and learn the new demapping functions, is required only when a (non-marginal) SNR working point change is identified. Alternatively, training for different SNR values can be performed offline or in a quasi-offline manner, with the obtained trained sets being stored in the receiver's memory. Another option is to provide the SNR level as an additional input to the LLRnet and train it within a wide range of SNRs. By doing so, the LLRnet can then learn to also generalize across SNR points.

Figs.~\ref{fig_LLR_QAM}(a)-(c) present, for 16-QAM, 64-QAM and 256-QAM (in which each symbol conveys $4$, $6$ and $8$ bits, respectively), in three different SNR levels, the explicit demapping functions of the real part of the (received) symbol into its carried odd-numbered bits' LLR for three different soft demodulation implementations: exact log-MAP~\eqref{eq_logMAP}, approximate max-log-MAP~\eqref{eq_maxlogMAP} and the proposed LLRnet. As expected, in the higher SNR regimes (lower row in each figure) for the examined three QAM modulations the max-log-MAP exhibits a good estimate of the optimal log-MAP rule. However, for the low and intermediate SNR regimes (two upper rows in each figure) the max-log-MAP only serves as a crude (low SNR) to reasonable (intermediate SNR) approximation to the exact demapper. On the other hand the soft bits inferred by the LLRnet practically coincide with the optimal ones in all three modulation cases across the entire relevant SNR range.

Next, the performance of LLRnet is evaluated in terms of measuring the PDSCH throughput. Figs.~\ref{fig_tput}(a)-(c) plot the PDSCH throughput as a function of SNR. The throughput is displayed both in terms of absolute values, in Mbps, and relative throughput, in percentage, w.r.t. the link's maximum possible throughput under the given configuration. It is evident that LLRnet (again, with only $K=8,16,32$ neurons populating the hidden layer, respectively) essentially provides the same throughput as the optimal log-MAP algorithm.\footnote{For high SNR points under $256$-QAM modulation, it was observed that using a hyperbolic tangent sigmoid transfer function yields slightly better performance than ReLU, at the expense of marginally more operations as depicted in Table~\ref{table_complexity}.} It is also observed that in the low and intermediate SNR levels the throughput corresponding to the tractable max-log-MAP substantially lags behind the throughput associated with the LLRnet (and log-MAP).

\begin{figure}[t!]
    \centering
    \includegraphics[width=\columnwidth, bb=0 0 420 315]{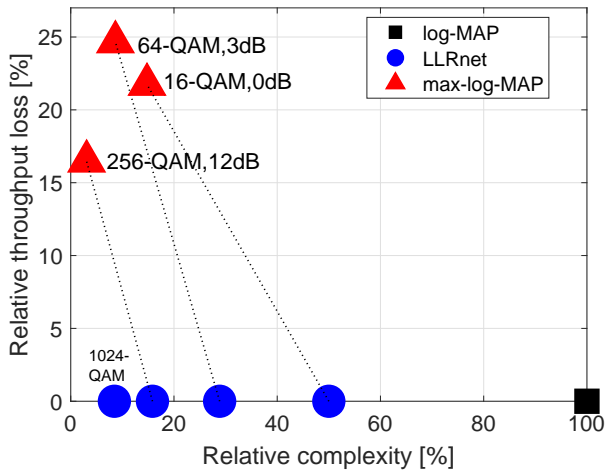}
    \caption{Comparison of LLRnet vs. max-log-MAP in a performance-complexity chart w.r.t. the exact log-MAP.}
    \label{fig_2D}
\end{figure}

The capability of LLRnet to successfully imitate the operation of the log-MAP algorithm is even more remarkable when comparing their computational complexity. Table~\ref{table_complexity} lists the number of (real) operations, per symbol, required for the execution of log-MAP and LLRnet for the three different QAM constellations. Note that without any loss in throughput, LLRnet costs about $50\%$, $70\%$, and $85\%$ \emph{less} total operations than the log-MAP algorithm. Bear in mind again these algorithms are run on symbol rate. For completeness, and although currently not yet being an integral part of 5G-NR, but of Wi-Fi 6 (802.11ax), the enumeration of the complexity savings (more than 90\%) of LLRnet vs. log-MAP for the 1024-QAM are listed in Table~\ref{table_complexity_1024}. The overhead complexity of the training phase itself in this application is relatively small since the learning relies on only few symbols within a single received slot, thus it is omitted from the comparison in Tables~\ref{table_complexity} and~\ref{table_complexity_1024}. In addition, the number of epochs required for training the LLRnet was typically below $30$. In this particular case of QAM constellations which can be divided into two separate PAM constellations, it is interesting to observe how LLRnet successfully learns, as anticipated, to null half of the hidden layer's weights. Alternatively in such a case, one can simply use two smaller LLRnet architectures processing separately the real and imaginary parts of the symbol.

Fig~\ref{fig_2D} visually compares the three different demapping algorithms through the construction of a two-dimensional chart in which the horizontal axis measures the relative complexity (w.r.t. the brute-force exact log-MAP, in percentage), and the vertical axis measures the relative loss in throughput (again, w.r.t. log-MAP, in percentage). For three different working points, with three different QAM modulation schemes, it is illustrated how the implementation of LLRnet, rather than the popular max-log-MAP, yields an improved position in the performance-complexity domain: Diminishing to practically zero with LLRnet, original throughput losses between $15\%$ to $25\%$, originated from the max-log-MAP, yet still maintaining feasible computational complexity burden via LLRnet. Observe how for each tested case, the position in the two-dimensional chart, based on max-log-MAP, moves into a new cost-effective position due to LLRnet (namely, {\color{red}$\blacktriangle$}, connected through a dashed line, shift into {\color{blue}$\bullet$}, located in a better position). Also marked are the performance-complexity metric position for the case of LLRnet in 1024-QAM and the reference metric position for the exact log-MAP (\ie, $\blacksquare$ at $100\%$ complexity and zero throughput loss point).

\subsection{Packet Error Rate in DVB-S.2}
\begin{figure}[t!]
    \centering
    \includegraphics[width=\columnwidth, bb=0 0 149 115]{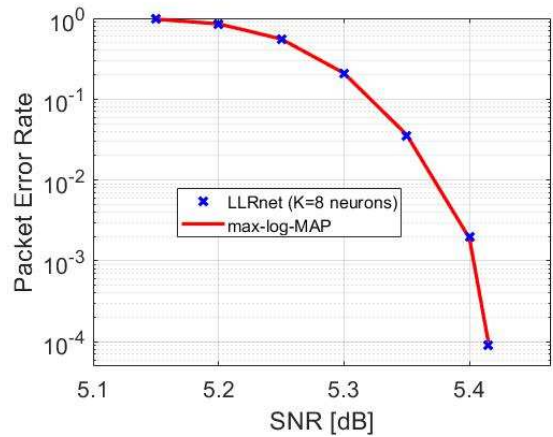}
    \caption{Packet error rate vs. SNR for DVB-S.2 link with modulation 8-PSK and LDPC code rate 3/5.}
    \label{fig_per_8psk}
\end{figure}

In this example, LLRnet is incorporated in the second generation digital video broadcasting standard (DVB-S.2) for broadband satellite communications. In this application example, LLRnet, with $K=8$ neurons in the hidden layer, is trained to mimic the operation of the approximate max-log-MAP algorithm in the demapping of 8-PSK modulation (thus $N=1$). A $3/5$ LDPC (decoded with a maximum of 50 iterations) and BCH codes serve as the inner and outer codes, respectively, in an AWGN channel. $1000$ DVB-S.2 frames were simulated for each SNR point. The training stage for this application example follows the procedure described for the training in Section~\ref{sec_5G}.

Fig.~\ref{fig_per_8psk} plots the simulated packet error rate as a function of the SNR using two implementations of soft demodulation: max-log-MAP and LLRnet, this time learned to imitate the approximate max-log-MAP (with $\sigma^{2}=\sigma_{0}^{2}$), rather than the exact log-MAP as in Section~\ref{sec_5G}. It can be observed that a perfect alignment in the performance curves of the two algorithms is achieved. However, it should be noted, that in this particular case, using LLRnet with $K=8$ neurons, the two implementations require roughly the same total number of operations (about 64 per symbol). It may happen that deep, rather than shallow, learning architectures could demonstrate, in addition to accurate inference of the LLRs, also further computational savings for 8-PSK and higher PSK, or amplitude and PSK (APSK) constellations.

\section{Conclusion}\label{sec_Conclusion}
In this paper, a "Machine-LLRning" approach is revealed for which a simple neural network architecture is trained to efficiently soft demodulate symbols to their bit LLRs, thus utilizing an artificial intelligence paradigm straight into one of the most generic building blocks of the physical layer processing. The proposed concept of LLRnet can be extended not only for multi-layer deep learning architectures, but also be integrated in a more holistic trainable receiver structure, jointly carrying out the tasks of neural demodulator (as proposed in this contribution), along with trainable quantization of the LLRs, and a trainable channel decoding. An evaluation of the LLRnet under realistic dynamical link adaptation, rather than a fixed modulation and coding scheme (MCS), is also called for. Furthermore, by feeding the channel estimation itself into the LLRnet, the tasks of MIMO and multiuser detection could also be potentially tackled in a straightforward manner in such a framework.

\bibliographystyle{IEEEtran}
\bibliography{IEEEabrv,machine_LLRning_Ori_19}

\end{document}